\documentstyle[aps,preprint]{revtex}
\tightenlines
\begin{document}
\draft
\title{ Propagators of light scalar mesons }
\author
{N.N. Achasov
\thanks{achasov@math.nsc.ru}
\ and A.V. Kiselev
\thanks{kiselev@math.nsc.ru}}

\address{
   Laboratory of Theoretical Physics,
 Sobolev Institute for Mathematics, Novosibirsk, 630090, Russia}

  \date{\today}
\maketitle

\begin{abstract}
For the first time, as far as we know, in field theory  are found
explicit forms of propagators satisfying the K\"allen -- Lehmann
representation. To be exact, it is shown that scalar meson
propagators, taking into account a virtual intermediate state
contribution to the resonance self-energy, satisfy the K\"allen --
Lehmann representation in the wide domain of coupling constants of
the light scalar mesons with the two-particle states.   It is
proposed to use these propagators in routine fitting data about
light scalar mesons to reveal physics underlying  the light scalar
mesons.
\end{abstract}

\vspace*{1cm}
 \pacs{ PACS number(s):  11.10.St, 11.55.Fv, 12.39.Fe}

Study of the nature of light scalar resonances has become a
central problem of nonperturbative QCD. The point is that the
elucidation of their nature is important for understanding both
the confinement physics and the chiral symmetry realization way in
the low energy region, i.e., the main consequences of QCD in the
hadron world. Actually, what kind of interaction at low energy is
the result of the confinement in the chiral limit? Is QCD
equivalent to the nonlinear $\sigma$ model or the linear one at
low energy?

The experimental nonet of the light scalar mesons \cite{pdg-2002},
the putative $f_0(600)$ (or $\sigma (600)$) and $\kappa (700-900)$
mesons and the well established $f_0(980)$ and $a_0(980)$ mesons
\cite{fourquarks}, suggests the  $U_L(3)\times U_R(3)$ linear
$\sigma$ model. History of the linear $\sigma$ model is rather
long, so that the list of its participants, quoted  in Ref.
\cite{gellman}, is far from complete. Hunting   the light $\sigma$
and $\kappa$ mesons had begun in the sixties already and a
preliminary information on the light scalar mesons in Particle
Data Group Reviews had appeared at that time. But long-standing
unsuccessful attempts to prove their existence in a conclusive way
entailed general disappointment and  information on these states
disappeared from Particle Data Group Reviews. One of principal
reasons against the $\sigma$ and $\kappa$ mesons was the fact that
both  $\pi\pi$ and $\pi\kappa$ scattering phase shifts do not pass
over $90^0$ at putative resonance masses. Situation changes when
it was shown \cite{annshgn-94} that in the linear $\sigma$ model
there is a negative background phase which hides the $\sigma$
meson. It has been made clear that shielding of wide lightest
scalar mesons in chiral dynamics is very natural. This idea was
picked up, see, for example, Ref. \cite{ishida}, and triggered new
wave of theoretical and experimental searches for the $\sigma$ and
$\kappa$ mesons, see Particle Data Group Review \cite{pdg-2002}.

In theory the principal problem is impossibility to use the linear
$\sigma$ model in the tree level approximation inserting widths
into $\sigma$ meson propagators   because such an approach breaks
the both unitarity and Adler self-consistency conditions
\cite{annshgn-94}. Strictly speaking, the comparison with the
experiment requires the nonperturbative calculation of the process
amplitudes \cite{nonperturbative}. Nevertheless, now there are the
possibilities to estimate odds of the $U_L(3)\times U_R(3)$ linear
$\sigma$ model to  the underlying  physics of the light scalar
mesons in phenomenology \cite{sigma}. Really, even now there is a
huge body of information about the $S$ waves of different
two-particle pseudoscalar states and what is more the relevant
information goes to press almost continuously from BES, BNL, CERN,
CESR, DA$\Phi$NE, FNAL, KEK, SLAC, and others. As for theory, we
know quite a lot about the scenario under discussion: the nine
scalar mesons, the putative chiral masking \cite{annshgn-94} of
the $\sigma (600)$ and $\kappa( 700-900)$ mesons, the unitarity
and Adler self-consistency conditions. In addition, there is the
light scalar meson treatment motivated by field theory. The
foundations of this approach were formulated in Refs.
\cite{adsh-79,adsh-80,adsh-1980,adsh-84}. In Refs.
\cite{adsh-79,adsh-80,adsh-1980} there were introduced into
practice the propagators of light scalar mesons $1/D_R(m^2)$
\cite{indirect}. The inverse propagator
\begin{equation}
\label{propagator}
 D_R(m^2) = m_R^2-m^2+ Re\left
(\Pi_R(m_R^2)\right )-\Pi_R(m^2)
\end{equation}
where $ Re\left (\Pi_R(m_R^2)\right )- \Pi_R(m^2)$
 takes into account the finite width
corrections of the light scalar $R$ resonance, which take into
account the contribution of  the two-particle virtual intermediate
$ab$ states to self-energy of the $R$ resonance,
\begin{equation}
\label{Pi}
 \Pi_R(m^2)=\sum_{ab}\Pi_R^{ab}(m^2)\,.
\end{equation}
In real axis of $m^2$
\begin{equation}
\label{im}
   Im\left (
\Pi_R(m^2)\right )=m\Gamma_R (m)=m\sum_{ab}\Gamma(R\to
 ab,m)\theta (m-m_a-m_b)
\end{equation}
where
\begin{equation}
\label{width}\Gamma(R\to ab,m)=\frac{g_{Rab}^2}{16\pi
m}\rho_{ab}\left(m^2\right)
\end{equation}
is the width of the $R\to ab$ decay, $m=m_{ab}$ is the invariant
mass of the $ab$ state, $g_{Rab}$ is the coupling constant of the
$R$ scalar resonance with the two particle $ab$ state
\cite{renormalization}, and
\begin{equation}
\label{rho-ab} \rho_{ab}\left(m^2\right)=\sqrt{\left
(1-\frac{m_+^2}{m^2}\right )\left (1-\frac{m_-^2}{m^2}\right
)}\,\,,\qquad m_{\pm}=m_a\pm m_b\,.
\end{equation}

 Below is shown that  propagators under discussion satisfy the
K\"allen -- Lehmann representation \cite{ab}
\begin{equation}
\label{k-l}
\frac{1}{D_R(m^2)}=\frac{1}{\pi}\int_{m_0^2}^{\infty}\frac{Im\left
(\frac{1}{D_R(\bar{m}^2)}\right )}{\bar{m}^2 - m^2 -
i\varepsilon}\,
d\bar{m}^2=\frac{1}{\pi}\int_{m_0^2}^{\infty}\frac{\bar{m}\Gamma_R(\bar{m})}{\left
|D_R(\bar{m}^2)\right |^2\left (\bar{m}^2 - m^2 -
i\varepsilon\right )}\, d\bar{m}^2
\end{equation}
in the wide domain of coupling constants of the scalar $R$
resonance with the two-particle $ab$ states, here
$m_0^2=(m_a+m_b)^2$ is the lowest threshold.

Recall that the one-loop contribution to the self-energy of the
$R$ resonance from the two-particle intermediate  $ab$ states
satisfies the dispersion relation with one subtraction. Let us
subtract at $m^2 = (m_a+m_b)^2$ \cite{subtraction}
\begin{equation}
\label{selfenergyab}
\Pi_R^{ab}(m^2)=\frac{1}{\pi}[m^2-(m_a+m_b)^2]\int_{(m_a+m_b)^2}^{\infty}\frac{\bar{m}\Gamma(R\to
ab,\bar{m})}{[\bar{m}^2-(m_a+m_b)^2](\bar{m}^2-m^2-i\varepsilon)}\,
d\bar{m}^2\,.
\end{equation}
When $m^2<(m_a+m_b)^2$
\begin{equation}
\label{negmon}
 \Pi_R^{ab}(m^2)< 0\,,\quad
\frac{d\Pi_R^{ab}(m^2)}{dm^2}=
\frac{1}{\pi}\int_{(m_a+m_b)^2}^\infty\frac{\bar{m}\Gamma (R\to
ab,\bar{m}) }{\left (\bar{m}^2 - m^2\right )^2}d\bar{m}^2 > 0\,.
\end{equation}

 Clearly, the K\"allen -- Lehmann representation takes place only if
\begin{equation}
\label{condition}
  D_R(z)=
m_R^2-z+Re\left (\Pi_R(m_R^2)\right ) - \Pi_R(z)\neq 0
\end{equation}
over the whole complex plane $m^2\equiv z=x+iy$.

Let us consider at first when $Im\left (D_R(x=m^2)\right )=0$. It
follows from Eqs. (\ref{propagator}), (\ref{Pi}), and
(\ref{selfenergyab}) that
\begin{eqnarray}
\label{imd}
 && Im\left (D_R(z)\right )=  -y\left (1 +
\sum_{ab}\frac{1}{\pi}\int_{(m_a+m_b)^2}^{\infty}\frac{\bar{m}\Gamma(R\to
ab, \bar{m})}{|\bar{m}^2- z|^2}\, d\bar{m}^2\right ).
\end{eqnarray}
Consequently, $Im\left (D_R(z)\right )=0$ only in real axis, when
$y=0$. In additional, see Eqs. (\ref{propagator}) and (\ref{im}),
$ Im\left (D_R(x=m^2)\right )= - m\Gamma_R(m)$. So, $Im\left
(D_R(x=m^2)\right )= 0$ only at $m^2 < m_0^2$.

Let us find now noughts of $Re\left (D_R(x=m^2)\right )=
D_R(x=m^2)$  at $m^2< m_0^2$
\begin{eqnarray}
\label{red0}
 && D_R(m^2)= m_R^2-m^2+Re\left
(\Pi_R(m_R^2)\right ) - \Pi_R(m^2)=0\,.
\end{eqnarray}

 The inverse
propagator $D_R(m^2)$  increases monotone when $m^2$ decreases at
$m^2 < m_0^2$   because
\begin{equation}
\label{monotondm0} \frac{dD_R(m^2)}{dm^2}= -1 -
\frac{d\Pi_R(m^2)}{dm^2} = - 1 -
\frac{1}{\pi}\int_{m_0^2}^\infty\frac{\bar{m}\Gamma_R(\bar{m})
}{\left (\bar{m}^2 - m^2\right )^2}d\bar{m}^2 < 0\,.
\end{equation}

So, the single nought of $D_R(m^2)$ is provided when
\begin{equation}
\label{nought} D_R(m_0^2)= m_R^2 -  m_0^2 + Re\left
(\Pi_R(m_R^2)\right ) - \Pi_R(m_0^2) \leq 0
\end{equation}
or
\begin{equation}
\label{nought1}
 m_R^2 - m_0^2   \leq  \Pi_R(m_0^2)- Re\left (\Pi_R(m_R^2)\right )=
\sum_{ab}\Pi_R^{ab}(m_0^2)- \sum_{ab}Re\left
(\Pi_R^{ab}(m_R^2)\right )\,.
\end{equation}

The left-hand side of Eq. (\ref{nought1}) is positive. As for the
right-hand side of  Eq. (\ref{nought1}), the contribution of the
every $ab$ channel, under the threshold of which the $R$ resonance
is, $m_R < m_a + m_b$\,,
\begin{equation}
\label{under}
\Pi_R^{ab}(m_0^2)- Re\left (\Pi_R^{ab}(m_R^2)\right
)= \Pi_R^{ab}(m_0^2)- \Pi_R^{ab}(m_R^2) < 0
\end{equation}
according to Eqs. (\ref{negmon}). So, only the $ab$ channels, the
thresholds of which are under the $R$ resonance, $m_a + m_b <
m_R$, can bring the threat to the K\"allen -- Lehmann
representation.

As for the light scalar meson case, $\sigma (600)$, $\kappa
(700-900)$, $f_0(980)$, $a_0(980)$, there is only one channel, the
threshold of which is under the $R$ resonance, $\pi\pi$, $\pi K$,
$\pi\pi$, $\pi\eta$ respectively. Let us consider at first the
one-channel case when the decay threshold is lower than the $R$
resonance, $m_a + m_b < m_R$. This scenario  is  most vulnerable
from the point of the K\"allen -- Lehmann representation view. One
has
\begin{eqnarray}
 \label{Piabhigher}
&& \Pi_R(m^2)= \Pi_R^{ab}(m^2)\equiv
\frac{g^2_{Rab}}{16\pi^2}P^{ab}(m^2)
=\frac{g^2_{Rab}}{16\pi^2}\left \{ \frac{\left ( m^2-m_+^2\right
)}{ m^2}\frac{m_-}{m_+}\ln \frac{m_a}{m_b} \right.
\nonumber\\[1pc] &&\left . + \rho_{ab}\left(m^2\right)\left [i\pi
+\ln\frac{\sqrt{m^2-m_-^2}-
\sqrt{m^2-m_+^2}}{\sqrt{m^2-m_-^2}+\sqrt{m^2-m_+^2}}\,\right
]\right\}
\end{eqnarray}
at $m \geq m_+ = m_a + m_b\,,$ here and hereafter $m_a\geq m_b\,.$
 As is evident from Eqs.
(\ref{nought1}) and  $\Pi^{ab}_R(m_0^2=m^2_+)=0$, the K\"allen --
Lehmann representation is valid if
\begin{eqnarray}
\label{k-l one}
 && m_R^2 - m_+^2 > -
\frac{g^2_{Rab}}{16\pi^2}Re\left (P^{ab}(m_R^2)\right )\,.
\end{eqnarray}

The one-channel propagators can be actual when treating the
$\sigma (600)$ resonance, the $\pi\pi$ channel, and the  $\kappa
(700-900)$ resonance, the $\pi K$ channel. Let us find what the
coupling constants are allowed from the K\"allen -- Lehmann
representation view in these cases.
\begin{eqnarray}
\label{gsigmapipigkappapik} && g^2_{\sigma\pi\pi}/16\pi^2< \left (
g^c_{\sigma\pi\pi}\right )^2/16\pi^2=
-\left(m_\sigma^2-4m_\pi^2\right)\Bigm/Re\left
(P^{\pi\pi}(m_\sigma^2)\right ) \approx
0.1\,\mbox{GeV}^2\,,\nonumber\\ && g^2_{\kappa\pi K}/16\pi^2<\left
( g^c_{\kappa\pi K}\right )^2/16\pi^2= -\left [m_\kappa^2 - \left
(m_\pi+ m_K\right )^2\right ]\Bigm/Re\left (P^{\pi\kappa}
(m_\kappa^2)\right ) \approx 0.4\,\mbox{GeV}^2\,,
\end{eqnarray}
where $g^2_{\sigma\pi\pi}= 1.5g^2_{\sigma\pi^+\pi^-}\,,$
$g^2_{\kappa\pi K}= 1.5g^2_{\kappa^+\pi^+ K^0}\,,$ $m_\kappa =0.8$
GeV. $\Gamma(\sigma\to\pi\pi,\,m_\sigma)<\Gamma_\sigma^c\approx
0.5$ GeV, $\Gamma(\kappa\to\pi K,\,m_\kappa
)<\Gamma_\kappa^c\approx 0.8$ GeV.

The two-channel propagators are often used when treating the
$f_0(980)$ resonance, the $\pi\pi$ and $K\bar K$ channels, and the
$a_0(980)$ resonance, the $\pi\eta$ and $K\bar K$ channels. To
consider this issue we need
\begin{equation}
\label{Piabbetween} \Pi^{ab}_{R}(m^2)=
\frac{g^2_{Rab}}{16\pi^2}\left[
\frac{\left(m^2-m_+^2\right)}{m^2}\frac{m_-}{m_+}\ln
\frac{m_a}{m_b}-
2|\rho_{ab}(m)|\arctan\frac{\sqrt{m^2-m_-^2}}{\sqrt{m_+^2-m^2}}\,
\right ]
\end{equation}
at $m_- = m_a - m_b\leq m\leq m_+ = m_a + m_b$.

As is easy to see from Eq. (\ref{nought1}), there are the $\left
(g_{f_0K\bar K}/g_{f_0\pi\pi}\right )^2$ and $\left (g_{a_0K\bar
K}/g_{a_0\pi\eta}\right )^2$ domains in the two-channel cases
where the right sides ($R=f_0$ or $R=a_0$) of Eq. (\ref{nought1})
are negative or vanish, in other words, the K\"allen -- Lehmann
representation holds for any $g_{f_0\pi\pi}^2$ and
$g_{a_0\pi\eta}^2$ respectively:
\begin{eqnarray}
\label{f0a0domain}
 && \left (g_{f_0K\bar K}/g_{f_0\pi\pi}\right
)^2\geq r_{f_0}^c =Re\left(P^{\pi\pi}(m^2_{f_0})\right
)\Biggm/\left
 [ P^{K\bar K}\left(4m_\pi^2\right )-P^{K\bar K}\left(m_{f_0}^2\right ) \right
 ]\approx 2.5\,,\nonumber\\
 &&\left (g_{a_0K\bar
K}/g_{a_0\pi\eta}\right )^2\geq r_{a_0}^c = Re\left (
P^{\pi\eta}(m^2_{a_0})\right )\Biggm/\left
 [ P^{K\bar K}\left(\left (m_\pi+m_\eta\right )^2\right)-P^{K\bar K}\left(m_{a_0}^2\right )\right
 ] \approx 0.8\,,
\end{eqnarray}
where  $g^2_{f_0\pi\pi}= 1.5g^2_{f_0\pi^+\pi^-}$, $g^2_{f_0K\bar
K}= 2g^2_{f_0K^+K^-}$\,, and $g^2_{a_0K\bar K}=
2g^2_{a_0^0K^+K^-}=g^2_{a_0^+K^+\bar K^0}$\,.

The ratios of the coupling constants, when the $f_0(980)$ and
$a_0(980)$ resonances were treated with the propagators under
discussion, satisfied the requirements (\ref{f0a0domain}).

As for outside of these domains, the K\"allen -- Lehmann
representation allows the following coupling constants :
\begin{eqnarray}
\label{gf0pipiga0pieta}
 &&
 \frac{g^2_{f_0\pi\pi}}{16\pi^2}< \frac{\left (g^c_{f_0\pi\pi}\right
 )^2}{16\pi^2}=\left(m_{f_0}^2-4m_\pi^2\right)\Biggm/
 \Bigg\{- Re\left (P^{\pi\pi}(m^2_{f_0})\right )\nonumber\\
 &&+\left
(g_{f_0K\bar K}/g_{f_0\pi\pi}\right  )^2\left
 [P^{K\bar K}\left(4m_\pi^2\right) - P^{K\bar K}\left(m_{f_0}^2\right)\right
 ]\Bigg\}\,,\nonumber\\
 && \frac{g^2_{a_0\pi\eta}}{16\pi^2}< \frac{\left (g^c_{a_0\pi\eta}\right )^2}{16\pi^2} = \left [m_{a_0}^2 - \left
(m_\pi+ m_\eta\right )^2\right ]\Biggm/ \Bigg\{- Re\left (
P^{\pi\eta}(m^2_{a_0})\right )\nonumber\\ &&+\left (g_{a_0K\bar
K}/g_{a_0\pi\eta}\right  )^2\left
 [P^{K\bar K}\left(\left (m_\pi+m_\eta\right )^2\right )
 - P^{K\bar K}\left(m_{a_0}^2\right)\right
 ]\Bigg\}\,.
\end{eqnarray}

 To experience what values of $g_{f_0\pi\pi}$ and
$g_{a_0\pi\eta}$ are allowed by the K\"allen -- Lehmann
representation when Eqs. (\ref{f0a0domain}) are not satisfied, we
consider a deliberately non-real case of the weak coupling of
$f_0(980)$ and $a_0(980)$ with the $K\bar K$ channel which gives a
very conservative estimate. Suppose $\left (g_{f_0K\bar
K}/g_{f_0\pi\pi}\right )^2=\left (g_{a_0K\bar
K}/g_{a_0\pi\eta}\right )^2=1/3$ then $\left (g_{f_0\pi\pi}\right
)^2/16\pi^2<\left (g^c_{f_0\pi\pi}\right )^2/16\pi^2\approx
0.3\,\mbox{GeV}^2$,
$\Gamma(f_0\to\pi\pi,\,m_{f_0})<\Gamma_{f_0}^c\approx 0.9$ GeV,
and $\left (g_{a_0\pi\eta}\right )^2/16\pi^2<\left
(g^c_{a_0\pi\eta}\right )^2/16\pi^2\approx 0.9\,\mbox{GeV}^2$,
$\Gamma(a_0\to\pi\eta,\,m_{a_0})<\Gamma_{a_0}^c\approx 1.8$ GeV
\cite{weak}.

Taking into account the $\eta\eta$, $\eta\eta^\prime$ and
$\pi\eta^\prime$ channels only extends the coupling constant
domains where the K\"allen -- Lehmann representation holds. The
inclusion of the $\eta\eta^\prime$ and $\pi\eta^\prime$ channels
requires
\begin{equation}
\label{lower} \Pi^{ab}_{R}(m^2)=
\frac{g^2_{Rab}}{16\pi^2}\left[\frac{\left ( m^2-m_+^2\right
)}{m^2}\frac{m_-}{m_+}\ln \frac{m_a}{m_b}-
\rho_{ab}(m)\ln\frac{\sqrt{m_+^2-m^2}-
\sqrt{m_-^2-m^2}}{\sqrt{m_+^2-m^2}+\sqrt{m_-^2-m^2}}\,\right ]
\end{equation}
 at $ m \leq m_- = m_a - m_b$ and $m^2\leq 0$.

The K\"allen -- Lehmann representation guarantees the unitarity
condition for the branching ratious
\begin{eqnarray}
\label{unitarity} && 1\equiv \sum_{ab}Br(R\to ab)=
\sum_{ab}\frac{1}{\pi}\int_{(m_a+m_b)^2}^{\infty}\frac{\bar{m}\Gamma(R\to
ab,\,\bar{m})}{\left |D_R(\bar{m}^2)\right |}\, d\bar{m}^2
\end{eqnarray}
which follows from Eqs. (\ref{k-l}), (\ref{propagator}),  and
(\ref{Piabhigher}) when $m^2\to\infty$ \cite{1980}.

 Recall that to
satisfy Eq. (\ref{unitarity}) the very popular Flatt\'e formulas
\cite{flatte} require a factor which considerably differs from 1,
see, for example, \cite{lenya}.  The Flatt\'e formulas ensue from
our ones by the following substitutions
\begin{eqnarray}
\label{flatte}
 && Re\left (\Pi_R^{ab}(m_R^2)\right ) - \Pi_R^{ab}(m^2)\to
 -i\frac{g^2_{Rab}}{16\pi}\rho_{ab}\left(m^2\right)\,,\quad  m_a +
 m_b\leq m\,; \nonumber\\
&& Re\left (\Pi_R^{ab}(m_R^2)\right ) - \Pi_R^{ab}(m^2)\to
 \frac{g^2_{Rab}}{16\pi}\left |\rho_{ab}\left(m^2\right)\right |\,,\quad m_a - m_b\leq m \leq  m_a
 + m_b\,.
\end{eqnarray}
As for $m\leq m_a-m_b$ and $m^2 < 0$, the analytic continuation of
the Flatt\'e formulas in this region has no physical sense. In
addition, the Flatt\'e formulas keep back traps for users. The
point is that  $Re\left ( D_R(m^2)\right ) $ has nought not at
$m_R^2$ but at the renormalized mass square $M_R^2=m_R^2 + \left
(g^2_{RK\bar K}/16\pi\right )\left |\rho_{K\bar K}\left
(M_R^2\right )\right |$ in the two channel $f_0(980)$ and
$a_0(980)$ treatment with the Flatt\'e formulas. Consequently, as
long as a user would like to associate the peak location in the
mass distribution
\begin{equation}
\label{spectrum}
\frac{dN_{ab}(m)}{dm}=N_{ab}\frac{2m^2}{\pi}\frac{\Gamma (R\to
ab,\, m)}{\left |D_R\left (m^2\right )\right |^2},\quad
R=f_0(980),\, ab=\pi\pi;\ R=a_0(980),\, ab=\pi\eta,
\end{equation}
with $m^2_R$, he will be in the region of the weak coupling of
$f_0(980)$ and $a_0(980)$ with the $K\bar K$ channel, a more
detailed consideration can be found in Ref. \cite{molecule}.

The propagators under discussion are used routinely by the Sobolev
Institute for Mathematics Group \cite{add}. They are used also by
the SND \cite{snd}, CMD-2 \cite{cmd2}, and KLOE \cite{kloe}
Collaborations in treating the $\phi\to\gamma
f_0(980)\to\gamma\pi\pi$ and $\phi\to\gamma
a_0(980)\to\gamma\pi\eta$ decays. In addition, these propagators
are used also in Refs. \cite{giulia}.  They have the ideal
properties, being only a little more complicated than the Flatt\'e
formulas \cite{byproduct}. We propose to use these propagators
routinely in treating the mass spectra $f_0(980)\to\pi\pi$ and
$a_0(980)\to\pi\eta$ to reveal physics underlying the light scalar
mesons.

Really, all information (coupling constants and masses) on scalar
mesons is extracted from spectra, but the study of spectra
requires knowledge of propagators. There are two means. 1) To use
the Flatt\'e formulas, which have Quantum Mechanics origin. These
formulas mean that virtual particles do not contribute to a
resonance self-energy, that is, these formulas mean that a
resonance is a weakly bound system. 2) To use our formulas, which
have Quantum Field Theory origin and take into account the
contribution of virtual particles to a resonance self-energy. Our
formulas are adequate to description of compact states ($q\bar q$,
$q^2\bar q^2$, and so on) strongly coupled with decay channels. In
addition, our formulas pass into the Flatt\'e ones at the limit of
weak coupling.

This work was supported in part by the RFBR Grant No. 02-02-16061
and the Presidential Grant No. 2339.2003.2 for support of Leading
Scientific Schools. A.V. Kiselev also thanks very much Dynasty
Foundation and ICFPM for scholarship.


\begin{thebibliography}{99}
\bibitem{nonperturbative}
A few approximate non-perturbative solutions in the frame of the
simplest linear $\sigma$-model are found in Ref.
\cite{annshgn-94}.
\bibitem{sigma}
 To be on the safe side,
notice that the linear $\sigma$ model does not contradict to
non-$q\bar q$ nature of the low lying scalars because Quantum
Fields can contain different virtual particles in different
regions of virtuality.
\bibitem{ab}
The $a$ and $b$ states are the stable pseudoscalar ones in all
specific cases which we consider, but all formulas are correct
also when the $a$ and $b$ states are the stable scalar ones. If
one considers the above $R$-resonance self-energy  as the one
chiral loop approximation in the linear $\sigma$ model, one would
expect that the higher order corrections reduce basically to the
coupling constant and mass renormalization. We mention that the
preliminary estimation in the two chiral loop approximation
supports these expectations. In other words, one would expect that
the formula in the one chiral loop approximation with the physical
coupling constants and mass is close to the exact formula.
\bibitem{subtraction}
Eq. (\ref{propagator}) does not depend on the subtraction of
$\Pi_R^{ab}\left (m^2\right )$ at any $m^2\leq \left
(m_a+m_b\right )^2$, or more generally, does not depend on an
addition of any real constant to $\Pi_R^{ab}\left (m^2\right )$.
\bibitem{byproduct}
The explicit analytic form of the new large family of the very
complicated integrals (\ref{k-l}) is found in our consideration as
a by-product.
\end{thebibliography}
\end{document}